\documentclass[journal,twoside,final]{IEEEtran}
\usepackage{amsmath}
\usepackage{amssymb}
\usepackage{amsthm}
\usepackage{bm}
\usepackage{comment}
\usepackage{float}
\usepackage{graphicx}
\usepackage{epstopdf}

\newtheorem{lemma}{Lemma}
\newtheorem{remark}{Remark}
\newtheorem{proposition}{Proposition}

\usepackage{cite}
\usepackage{enumitem}

\usepackage{xcolor}
\usepackage{lipsum}
\usepackage[ruled,norelsize]{algorithm2e}
\usepackage{algorithmic}

\allowdisplaybreaks

\begin{document}
%
%
\title{Beamforming Design for Pinching Antenna Systems with Multiple Receive Antennas}

\author{Enzhi Zhou, Yue Xiao, Ziyue Liu, Sotiris A. Tegos,~\IEEEmembership{Senior Member,~IEEE,} \\ Panagiotis D. Diamantoulakis,~\IEEEmembership{Senior Member,~IEEE,} George K. Karagiannidis,~\IEEEmembership{Fellow,~IEEE}
\vspace{-0.5 in}


\thanks{Enzhi Zhou is with the School of Computer and Software Engineering, Xihua University, Chengdu 610039, China (e-mail:ezzhou@mail.xhu.edu.cn).}
\thanks{
Y Xiao is with the School of Information Science and Technology, Southwest Jiaotong University, Chengdu 610031, China (e-mail: alice\_xiaoyue@hotmail.com. } 
\thanks{Ziyue Liu is with the School of Aeronautics and Astronautics, Xihua University, Chengdu 610039, China (e-mail:liuziyue\_2006@126.com).}
\thanks{
S. A. Tegos, P. D. Diamantoulakis and G. K. Karagiannidis are with the Department of Electrical and Computer Engineering, Aristotle University of Thessaloniki, 54124 Thessaloniki, Greece (e-mail: tegosoti@auth.gr, padiaman@auth.gr, geokarag@auth.gr).}
}
\setlength{\textfloatsep}{3pt }  
\maketitle

\begin{abstract}
Next-generation networks require intelligent and robust channel conditions to support ultra-high data rates, seamless connectivity, and large-scale device deployments in dynamic environments. 
While flexible antenna technologies such as fluid and movable antennas offer some degree of adaptability, their limited reconfiguration range and structural rigidity reduce their effectiveness in restoring line-of-sight (LoS) links. As a complementary solution, pinching antenna systems (PASs) enable fine-grained, hardware-free control of radiation locations along a waveguide, offering enhanced flexibility in challenging propagation environments, especially under non-LoS (NLoS) conditions.
This paper introduces a general and novel modeling framework for downlink PASs targeting users equipped with multiple receive antennas, addressing a practical yet underexplored scenario in the existing literature.
Specifically, we first derive an analytical relationship between the received signal-to-noise ratio and the pinching antenna (PA) positions, and based on this, we propose a two-layer placement strategy. First, we optimize the central radiation point using large-scale channel characteristics, and then we use a heuristic compressed placement algorithm to approximate phase alignment across multiple receive antennas and select a spatially compact set of active elements. Simulation results demonstrate notable performance gains over conventional single-antenna schemes, particularly in short-range scenarios with dense PAs and widely spaced user antennas.
\end{abstract}

\begin{IEEEkeywords}
Pinching antennas (PAs), beamforming, multiple antennas,  antenna positions.
\end{IEEEkeywords}


\vspace{-0.1 in}
\section{Introduction}

To realize the vision of next-generation communication networks, which aim to provide ultra-high data rates, seamless connectivity, and massive device deployments in dynamic environments, it is essential to ensure intelligent and robust wireless channels. However, the quality of wireless links is fundamentally constrained by the propagation environment, including small-scale fading and large-scale path loss.
To address these challenges, adaptive antenna architectures are needed to improve link robustness. Recent advances in flexible antennas, such as fluid and movable designs, enable limited spatial reconfiguration to accommodate user mobility and multipath effects. While effective in moderately dynamic scenarios, their performance declines under severe line-of-sight (LoS) blockage, especially at high frequencies, due to restricted reconfiguration range and structural rigidity.

Recently, pinching antennas (PAs) have emerged as a promising complementary solution for improving link reliability in non-LoS (NLoS) conditions. First introduced by DOCOMO in 2022, these antennas use small dielectric particles to activate radiation points along a waveguide.
By supporting flexible deployment and reconfigurable LoS connectivity, PAs provide a scalable and cost-effective enhancement, especially in obstructed environments, and represent a strong candidate for future communication networks.
A key challenge in designing of pinching antenna systems (PASs) is the joint optimization of antenna activation positions and beamforming. Most existing studies focus on downlink scenarios with single-antenna users, where the goal is to align the phases of signals from active elements to maximize coherent combining at the receiver \cite{lv2025beamtrainingpinchingantennasystems, 10896748}. However, when the user is equipped with multiple antennas, achieving simultaneous phase alignment across all receive antennas becomes significantly more complex. This introduces new challenges in PA position optimization that remain largely unaddressed in the current literature.

In \cite{ding25flexible}, the authors evaluated the path loss reduction and steering flexibility enabled by adaptive PA positioning. In \cite{tyr}, closed-form expressions for the coverage probability of randomly distributed users were derived, taking waveguide losses into account. In multi-user scenarios, the authors of \cite{sotiris_panos} introduced a scheduling strategy aimed at balancing user rates, while \cite{ding2025blockage} demonstrated that specific LoS blockages can enhance sum throughput by reducing user correlation. Furthermore, to mitigate intersymbol interference, in \cite{oikonomou2025ofdma}, an orthogonal frequency-division multiple access-based framework was introduced, and a max-min resource allocation problem was formulated to ensure user fairness.

To address the modeling gap in a downlink PAS with a multi-antenna user, this paper first establishes a mathematical relationship between the received signal-to-noise ratio~(SNR) and the transmit antenna positions. Based on this formulation, the central PAs is optimized using large-scale channel characteristics achieving coarse alignment with the user antenna layout.
A heuristic compressed antenna placement algorithm is then proposed. In each iteration, an approximate linear model of channel phase variation is derived based on the in-phase signal superposition condition, enabling the identification of near-optimal transmit positions for each user antenna. To enhance spatial efficiency, a minimum-range algorithm is applied to extract a concentrated subset of transmit positions, yielding an approximately optimal configuration.
Simulation results demonstrate that the proposed algorithm outperforms conventional schemes designed for single-antenna users. This improvement is especially evident in short-range scenarios involving dense PAs, a small number of receive antennas, and large inter-antenna spacing.

\vspace{-0.1 in}
\section{System Model and Problem Formulation}
\begin{figure}
	\vspace{-0.1cm}
	\centering
	\includegraphics[width=0.36\textwidth]{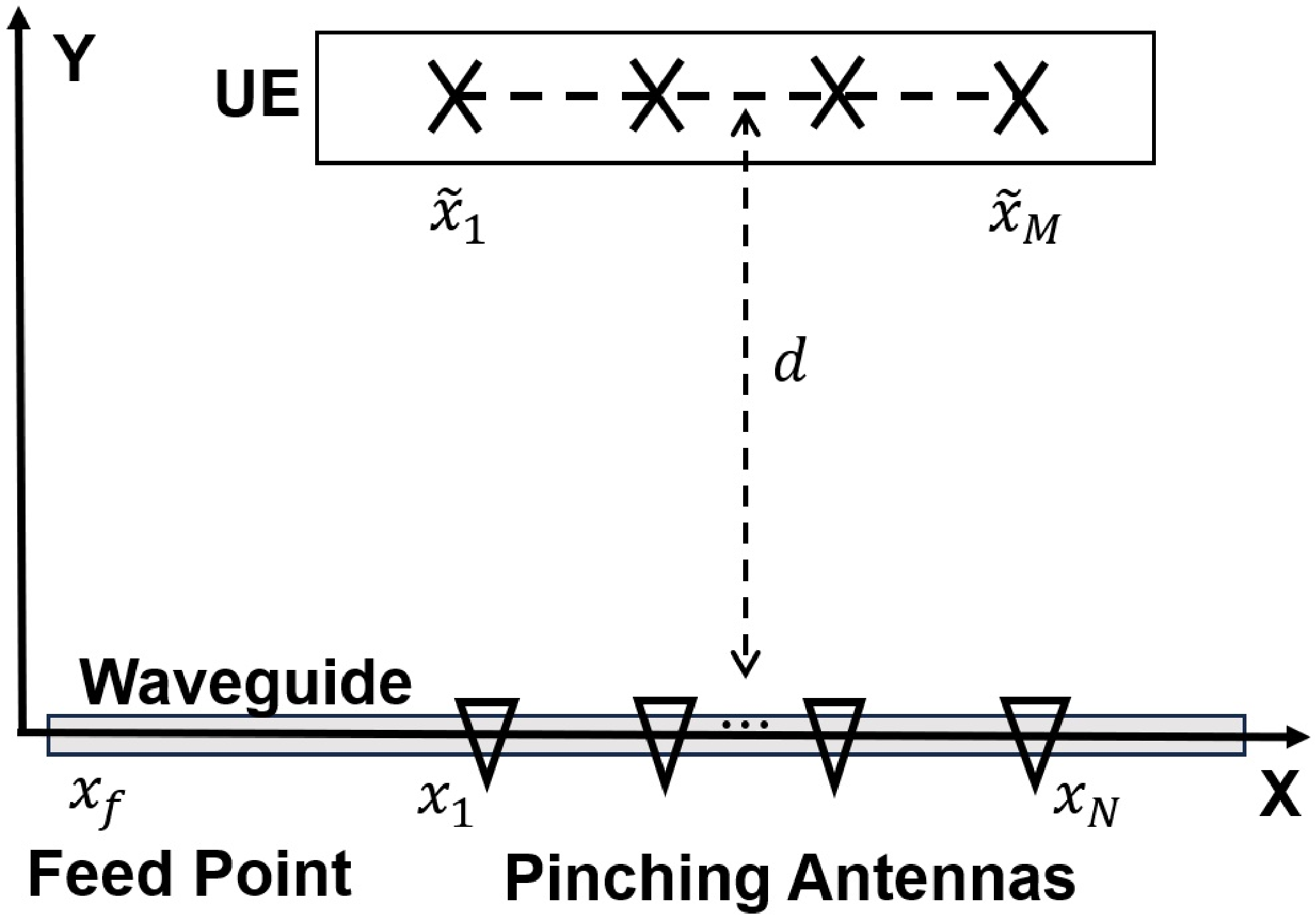}
	\vspace{-0.2cm}
	\caption{A downlink PAS with multiple receive antennas at the user.
}\label{fig:system_model}
\end{figure}

As shown in Fig. \ref{fig:system_model}, the system is modeled in a 2D coordinate plane, where a waveguide with $N$ uniformly spaced PAs lies along the x-axis. 
{\color{black} Since a single user is considered, the 2D model is equivalent to the 3D case in terms of relative geometry, and is adopted here for notational and analytical simplicity.}
A user equipped with $M$ antennas is positioned at a vertical distance $d$ from the waveguide. 
Without loss of generality, the coordinates of the $n$-th PA and the $m$-th user antenna are denoted by $\bm{\varphi}_n = (x_n, 0)$ and $\bm{u}_m = (\tilde{x}_m, d)$, respectively.
The signal feed point is $\bm{\varphi}_f = (x_f, 0)$, with $x_n > x_f$ and indices increasing from left to right. The waveguide endpoint is $\bm{\varphi}_e = (x_e, 0)$.

\setlength{\textfloatsep}{5pt } 

Based on the predefined coordinate system, the equivalent channel between the signal feed point and the $m$-th user antenna through the $n$-th PA is expressed as
\begin{equation}
h_{m,n} = \frac{{{\eta ^{\frac{1}{2}}}}}{\left\| \bm{\varphi}_n-\bm{u}_m\right\|}{e^{ - j \theta_{m,n}}},
\end{equation}
where $\eta = \frac{c^2}{16\pi^2 f_c^2}$, with $c$ denoting the speed of light and $f_c$ representing the carrier frequency.
Additionally, the channel phase is
\begin{equation}\label{eq:channel_phase_delay}
    \theta_{m,n} =\left( {\frac{{2\pi }}{\lambda }\sqrt {{{\left( {{{\tilde x}_m} - {x_n}} \right)}^2} + {d^2}}  + \frac{{2\pi }}{{{\lambda _g}}}\left( {{x_n} - {x_f}} \right)} \right), 
\end{equation}
where $\lambda$ is the free-space wavelength, and $\lambda_g = \lambda / n_{\mathrm{eff}}$ is the guided wavelength determined by the effective refractive index of the waveguide $n_{\mathrm{eff}}$.

Assuming negligible transmission loss in the waveguide and uniform power allocation across $N$ PAs, the received signal at the $m$-th user antenna is given as
\begin{equation}
y_m   = \sqrt {\frac{P}{N}} \sum\limits_{n = 1}^N {{h_{m,n}}{s}}  + {w_m},
\end{equation}
where $s$ is the normalized transmit symbol, $P$ is the total transmit power, and $w_m \sim \mathcal{CN}(0, \sigma_n^2)$ denotes the additive complex Gaussian noise at the $m$-th receive antenna with variance $\sigma_n^2$.
Let $h_m = \sum\nolimits_{n = 1}^N h_{m,n}$ denote the aggregated channel from all $N$ PAs to the $m$-th receive antenna. The overall channel vector is then given by $\mathbf{h} = [h_1, \ldots, h_M]$. At the user side, maximum ratio combining is employed as the optimal receive strategy, resulting in
\begin{equation}
    y = \sqrt {\frac{P}{N}} \sum\limits_{m = 1}^M {\frac{{h_m^*}}{{\left\| {\bf{h}} \right\|}}{h_m}s}  + \tilde w = \sqrt {\frac{P}{N}} {\left\| {\bf{h}} \right\|s}  + \tilde w,
\end{equation}
where $\tilde{w} = \sum\nolimits_{m = 1}^M \frac{h_m^*}{\|\mathbf{h}\|} w_m$ represents the post-processing noise, which follows a complex Gaussian distribution, i.e., $\tilde{w} \sim \mathcal{CN}(0, \sigma_n^2)$. 
Let $\tilde{\gamma} = \frac{P}{\sigma_n^2}$ denote the transmit SNR, the effective received SNR is given by $\gamma = \|\mathbf{h}\|^2 \tilde{\gamma}$, and the corresponding achievable rate is $\log_2(1 + \gamma)$. This expression indicates that maximizing the received SNR is equivalent to maximizing $\|\mathbf{h}\|^2$. Therefore, the SNR maximization problem is formulated as
\begin{subequations}\label{Problem_main}
\begin{align}
(\text{P}1):~ \mathop {\max }\limits_{{x_n}}  &\sum\limits_{m = 1}^M {{{\left| {\sum\limits_{n = 1}^N {{h_{m,n}}} } \right|}^2}}  \label{eq:objective_func} \\
 ~~ \text{s.t.}   ~~ & x_{i+1} - x_i \geq \frac{\lambda}{2}, \quad \forall i \in \{1{:}N-1\}
\label{eq:restrict_interval}  \\
&x_n \in [x_f, x_e], \quad \forall n \in \{1{:}N \}. \label{eq:restrict_left_end}\end{align}
\end{subequations} 
The constraint in \eqref{eq:restrict_interval} enforces a minimum spacing of $\lambda/2$ to avoid mutual coupling, while  \eqref{eq:restrict_left_end} ensures antenna placement within the waveguide. For simplicity, we assume a sufficiently long waveguide, boundary constraints are omitted in the following analysis \cite{pozar2012microwave}. 
To address the non-convex antenna placement problem (P1), we develop a greedy algorithm that sequentially selects antenna positions to maximize the received SNR, as detailed in the following section.

\vspace{-0.2 cm}
\section{Proposed Strategy for PA Placement}
To improve the received SNR while maintaining practical deployment simplicity, we propose a two-step antenna placement strategy, {\color{black} inspired by the approach presented in \cite{10896748}. First, when the center of PA array is positioned aligned with the user's x-coordinate,  the large-scale path loss is minimized.} 
Then, the remaining antennas are sequentially deployed outward from the central PA to ensure minimum spacing constraints are always met. 
In general, the antenna placement follows the order as
\begin{equation}\label{eq:antenna_placement_order}
\left[ \pi(1), \ldots, \pi(N) \right] = \left[ \tfrac{N}{2}, \tfrac{N}{2} + 1, \ldots, N, \tfrac{N}{2} - 1, \ldots, 1 \right],
\end{equation}
where $\pi(k)$ denotes the index of the $k$-th deployed antenna.
{\color{black}
Note that, although \eqref{eq:antenna_placement_order} is written in an even-$N$ form for simplicity, the center-outward symmetric deployment applies to both even and odd cases. 
For even $N$, the center is not unique, but either of the two middle antennas can be chosen without loss of generality. Moreover, since practical systems typically use even numbers of antennas, the even-$N$ formulation provides a representative modeling approach.
}
The placement of the first antenna is primarily based on large-scale fading. 
\color{black} As concluded in \cite{10896748}, the center position of PA array determines the overall large-scale fading.
To minimize large-scale path loss across all $M$ receive antennas, {\color{black} the optimal x-coordinate of the array center is determined by}
\begin{equation}\label{eq:path_loss_objective_function}
    \mathop {\max }\limits_x \;\; L\left(x\right)=\sum\limits_{m = 1}^M {\frac{1}{{{{\left( {x - {{\tilde x}_m}} \right)}^2} + {d^2}}}}, 
\end{equation}
where $L(x)$ represents the aggregated inverse path loss from the central antenna to all receive antennas.

\begin{lemma}
$L(x)$ has the following properties:
\begin{enumerate}
\item The global maximum of $L(x)$ is attained in the closed interval $[\tilde{x}_1, \tilde{x}_M]$.
\item If $\tilde{x}_M - \tilde{x}_1 < \frac{d}{\sqrt{3}}$, $L(x)$ is strictly concave over $[\tilde{x}_1, \tilde{x}_M]$ and the optimal solution is unique.
\end{enumerate}
\end{lemma}

\begin{IEEEproof}
We first compute the derivative of $L(x)$ as
\begin{equation}\label{eq:path_loss_derivation}
    L'\left(x\right) = \sum\limits_{m = 1}^M {\frac{{2\left( {{{\tilde x}_m} - x} \right)}}{{{{\left( {{{\left( {x - {{\tilde x}_m}} \right)}^2} + {d^2}} \right)}^2}}}}.
\end{equation}
It follows that $L'(x) > 0$ for $x < \tilde{x}_1$ and $L'(x) < 0$ for $x > \tilde{x}_M$, indicating that $L(x)$ is strictly increasing on $(-\infty, \tilde{x}_1)$ and strictly decreasing on $(\tilde{x}_M, +\infty)$. Since $L(x)$ is continuous on $[\tilde{x}_1, \tilde{x}_M]$, the global maximum is attained in this interval.
Next, the second derivative is
\begin{equation}
    L''\left(x\right) = \sum\limits_{m = 1}^M {\frac{{6{{\left( {x - {{\tilde x}_m}} \right)}^2} - 2{d^2}}}{{{{\left( {{{\left( {x - {{\tilde x}_m}} \right)}^2} + {d^2}} \right)}^3}}}}.
\end{equation}
It can be observed that if $|x - \tilde{x}_m| < \frac{d}{\sqrt{3}}$ for all $m$, then the numerator of each term is lower than 0, implying $L''(x) < 0$. This confirms that $L(x)$ is strictly concave over $[\tilde{x}_1, \tilde{x}_M]$, ensuring a unique optimal solution.
\end{IEEEproof}

\begin{lemma}\label{remark:symetric_receive_antenna}
In the special case where the antennas at the user side are symmetrically deployed around the point $\tilde x_c = 0.5(\tilde{x}_1 + \tilde{x}_M)$, {\color{black} the optimal x-coordinate of the PA array that maximizes the path gain is} $x = \tilde x_c$.
 \end{lemma}
\begin{IEEEproof}
Utilizing the symmetry, we have $\tilde x_{M-i+1}  = 2 \tilde x_c-\tilde x_i$. After some simple manipulations and by substituting $x = \tilde x_c$ into \eqref{eq:path_loss_derivation}, we have 
\begin{equation}\label{symmetry}
    {\frac{{2\left( {{{\tilde x}_i} - \tilde x_c} \right)}}{{{{\left( {{{\left( {\tilde x_c - {{\tilde x}_i}} \right)}^2} + {d^2}} \right)}^2}}}}+\frac{{2\left( {\left(2\tilde x_c-\tilde x_i \right) - \tilde x_c} \right)}}{{{{\left( {{{\left( {\tilde x_c - \left(2\tilde x_c-\tilde x_i \right)} \right)}^2} + {d^2}} \right)}^2}}}=0.
\end{equation}
By pairing symmetric indices $i = 1,\ldots, \frac{N}{2}$ according to \eqref{symmetry}, it follows that each pair of terms in \eqref{eq:path_loss_derivation} cancels out, leading to $L'(\tilde x_c) = 0$. Therefore, $x = \tilde x_c$ is a stationary point of \eqref{eq:path_loss_objective_function} and thus the optimal solution, which completes the proof.
\end{IEEEproof}

For symmetric receive antenna arrays, the center of the PA array should be optimally placed at $\tilde x_c$. Since the PA spacing is small, we set $ x_{\frac{N}{2}} = \tilde x_c$  for simplicity expression, and $\tilde x_c$ can be numerically optimized over $\left[\tilde x_1,\tilde x_M \right]$ via binary search or gradient descent.

After positioning the $\frac{N}{2}$-th PA, the locations of the remaining PAs, i.e., $\{\pi(2),\ldots,\pi(N)\}$ as defined in \eqref{eq:antenna_placement_order}, are optimized sequentially in a center-outward manner.
\subsubsection{Optimization Objective}
 Based on the deployment sequence defined in \eqref{eq:antenna_placement_order}, the cumulative channel at the $m$-th user antenna after placing the first $k-1$ PAs is 
 \begin{equation}
\hat h_{m,k-1} = \sum_{i=1}^{k-1} h_{m,\pi(i)} = \kappa_{m,k-1} e^{-j \Theta_{m,k-1}},
\end{equation}
where $\kappa_{m,k-1} = l_m \left| \sum_{i=1}^{k-1} e^{-j \theta_{m,\pi(i)}} \right|$ is the resulting channel amplitude, with the path loss from all PAs to the $m$-th user antenna being approximated as $l_m =  \sqrt{\eta /[ ( \tilde x_m - x_{\frac{N}{2}} )^2 + d^2  ]}  $, and 
$\Theta_{m,n}$ denotes the equivalent phase delay of the aggregated channel. 
Then, placing the $\pi(k)$-th PA induces a new signal component, and the updated channel becomes
\begin{equation}
\hat h_{m,k} = \hat h_{m,k-1} + l_m e^{-j \theta_{m,\pi(k)}}.
\end{equation}
Specifically, $\theta_{m,\pi(k)}$ denotes the phase delay caused by the $\pi(k)$-th PA at the $m$-th user antenna, which depends on its position $x_{\pi(k)}$ as defined in \eqref{eq:channel_phase_delay}. Accordingly, the corresponding channel gain is
\begin{equation}
    G_{k} = \sum\limits_{m=1}^M g_{m, k}=\sum\limits_{m=1}^M \left| \hat h_{m,k} \right| ^2,
\end{equation}
with $g_{m,k}$ being the channel gain of the $m$-th receive antenna. 
Until now, the position $x_{\pi(k)}$ is selected to locally maximize the channel gain $G_{k}$, i.e., 
\begin{equation}
x^o_{\pi(k)}=\arg\max_{x_{\pi(k)}} G_{k}.
\end{equation}

Moreover, because the total channel gain $G_{k}$ is a non-concave function of the antenna position $x_{\pi(k)}$ and may have multiple local maxima, it is difficult to optimize directly. 
To simplify the optimization, we analyze each per-antenna gain $g_{m,k}$ separately.
For the $m$-th receive antenna, the optimal position $x^o_{\pi(k)}$ of the newly added PA should produce a signal with a phase that aligns with the accumulated channel,
which is equivalent to 
\begin{equation}\label{eq:phase_align_condition}
  \Delta \theta_{m,\pi(k)} = \theta_{m,\pi(k)} ( {x_{\pi(k)}^o} ) - {\tilde \Theta _{m,k-1}} = 2 z\pi , ~z \in \mathbb{Z},
\end{equation}
where $\tilde \Theta_{m,k-1} \triangleq \bmod\left( \Theta_{m,k-1},\ 2\pi \right)$ is the normalized phase delay of the accumulated channel.
The periodic nature of the phase implies that the alignment constraint in \eqref{eq:phase_align_condition} admits multiple valid solutions, parameterized by the integer variable $z$.
Considering the $\lambda/2$ spacing constraint between adjacent antennas as specified in \eqref{eq:restrict_interval}, the left boundary of the feasible region to place the $n$-th PA is given by 
\begin{equation}\label{reference_region}
    x_n^s = {x_{n - 1}} + \operatorname{sgn} \left( {n - \frac{N}{2}} \right)\frac{\lambda }{2},n = 1, \ldots ,N,n \ne \frac{N}{2},
\end{equation}
where $\operatorname{sgn}(\cdot)$ denotes the sign function, which returns $+1$ if its argument is positive, $-1$ if negative, and $0$ otherwise.

\subsubsection{Configuration Design}
Based on the defined feasible region, the optimal PA position $x^o_{\pi(k)}$ corresponding to the user can be reformulated into finding the per-antenna candidate positions $x^o_{m,\pi(k),|z|}$ for all receive antennas.
The following result gives the closed-form expression for the optimal positions corresponding to the $m$-th receive antenna.

\begin{proposition}
The optimal PA positions corresponding to the $m$-th receive antenna  can be approximated by
\begin{equation}\label{eq:optimal_position_each_antenna}
    x^o_{m,\pi(k), \left| {z} \right|} \approx \frac{{\tilde \Theta _{m,k-1}} +  2 z \pi-  \theta_{m,\pi(k)}^s}{\theta_{m,\pi(k)}'} +  x_{\pi(k)}^s, ~ z\in \mathbb{Z},
\end{equation}
which corresponds to multiple candidate solutions due to the periodicity of the phase alignment condition. Here, $x_{\pi(k)}^s$ is given by \eqref{reference_region}.

\end{proposition}
\begin{IEEEproof}
The phase delay between the $\pi(k)$-th PA and the $m$-th user antenna can be approximated via a first-order Taylor expansion at the reference point $x_{\pi(k)}^s$ as
\begin{equation}\label{eq:single_antenna_phase_delay_approx}
\theta_{m,\pi(k)}(x_{\pi(k)}) \approx \theta_{m,\pi(k)}^s + \theta_{m,\pi(k)}' (x_{\pi(k)} - x_{\pi(k)}^s),
\end{equation}
where $\theta_{m,\pi(k)}^s$ is the channel phase evaluated at $x_{\pi(k)}^s$. According to \eqref{eq:channel_phase_delay}, the derivative of $\theta_{m,\pi(k)}$ is given by
\begin{equation}
\begin{aligned}
\theta_{m,\pi(k)}' &= \frac{2\pi}{\lambda}  \frac{x_{\pi(k)} - \tilde{x}_m}{\sqrt{(\tilde{x}_m-x_{\pi(k)} )^2 + d^2}} + \frac{2\pi}{\lambda_g}\\
&\overset{(a)}{\approx} \frac{2\pi}{\lambda} \frac{(x_{\pi(k)}^s - \tilde{x}_m)}{d} + \frac{2\pi}{\lambda_g}.
\end{aligned}
\end{equation}
Step (a) follows from the fact that the PA is centered within the user antenna span, thus the lateral spacing is negligible compared to the vertical distance $d$, and $x_{\pi(k)}$ varies minimally. Therefore, the derivative is approximated at the reference position $x_{\pi(k)}^s$. By substituting \eqref{eq:single_antenna_phase_delay_approx} into the optimal condition \eqref{eq:phase_align_condition}, we then obtain \eqref{eq:optimal_position_each_antenna}. Additionally, note that $z > 0$ when $\pi(k)>\frac{N}{2}$ and $z < 0 $ when $\pi(k)<\frac{N}{2}$, thus the index of the optimal antenna position is set as $\left|z\right|$.
\end{IEEEproof}

\begin{remark}
{\color{black} For the case of a single receive antenna, a simple
antenna placement scheme results from \eqref{eq:optimal_position_each_antenna}
by setting  by setting $z = 1$ and $z = -1$ for $\pi(k)>\frac{N}{2}$ and $\pi(k)<\frac{N}{2}$, respectively. This
scheme can achieve near-optimal performance while avoiding
the search-based algorithm as \cite{10896748}.}
\end{remark}

\begin{figure}
	\vspace{-0.1cm}
	\centering
	\includegraphics[width=0.49\textwidth]{
    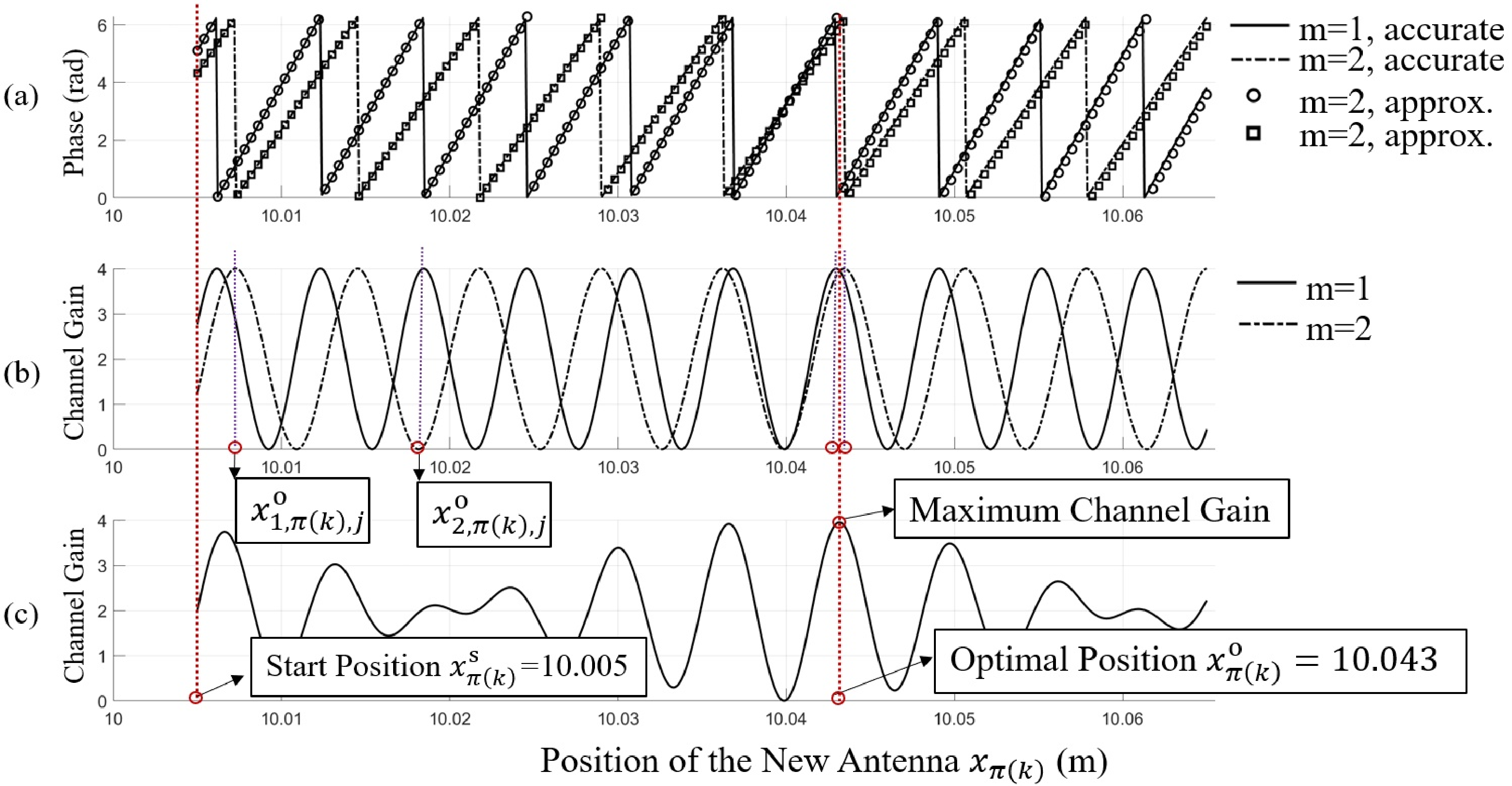}
    \vspace{-6mm}
	\caption{Impact of different antenna position on the phase difference and channel gain: (a) approximated phase difference {$ \tilde \Delta \theta_{m,\pi(k)}$}; (b) channel gain at each receive antenna {$g_{m,k}$}; (c) total aggregated channel gain  {$G_{k}$}.}\label{fig:effect_of_adding_antenna}
    \vspace{-0.1cm}
\end{figure}

To extend the analysis to systems with multiple receive antennas, we examine how the addition of a new PA influences the overall channel gain. Fig.~\ref{fig:effect_of_adding_antenna} illustrates a case with \(M = 2\) and \(N = 8\), where the $5$-th PA is placed given that the central antenna is fixed at \(x_4 = 10\,\text{m}\). The figure comprises three subplots: the phase difference \(\tilde{\Delta} \theta_{m,\pi(k)} = \bmod\left(\Delta \theta_{m,\pi(k)}, 2\pi\right)\) between the newly induced and accumulated channels at each receive antenna, the individual channel gain \(g_{m,k}\), and the total channel gain \(G_k\) as a function of the candidate position \(x_{\pi(k)}\).

Fig. \ref{fig:effect_of_adding_antenna} (a) shows that the phase difference increases with $x_{\pi(k)}$ and phase alignment occurs when $\Delta\theta_{m,\pi(k)}$ is an integer multiple of $2\pi$. Moreover, the accuracy of the linear approximation is confirmed, consistent with the result presented in Remark~1.
Additionally, Fig.~\ref{fig:effect_of_adding_antenna}(b) reveals that each receive antenna exhibits multiple local optimal positions $x^o_{m,\pi(k),j}$ due to the periodic nature of phase alignment, resulting in multiple peaks in the corresponding channel gain $g_{m,k}$.
Finally, Fig.~\ref{fig:effect_of_adding_antenna}(c) illustrates that the total channel gain $G_{k}$ exhibits non-monotonic behavior due to phase misalignment. The optimal antenna position lies near the centroid of \( \{x^o_{m,\pi(k),j}\}_{m=1}^M \).  When these positions are tightly clustered, the signals are almost phase-aligned, yielding peak gain.
 In this example, the maximum occurs at \( x = 10.043\,\text{m} \), where the optimal positions for all receive antennas nearly coincide.
\begin{remark}\label{remark:design_motivation}
Due to the periodic and antenna‐dependent nature of optimal positions, the joint alignment of phases between multiple reception antennas is challenging. To address this, we exploit the clustered‐position phenomenon in Fig. \ref{fig:effect_of_adding_antenna}. For each receive antenna $m$, multiple candidates $x^o_{m,\pi(k),|z|}$ arise from different phase indices $z$, while only one global position $x^o_{\pi(k)}$ is chosen for placement. These candidates cluster near the global optimum, enabling phase alignment and motivating the proposed algorithm.
\end{remark}

\begin{figure*}
    \centering
    \begin{minipage}{0.32\textwidth}
        \includegraphics[width=0.98\linewidth]{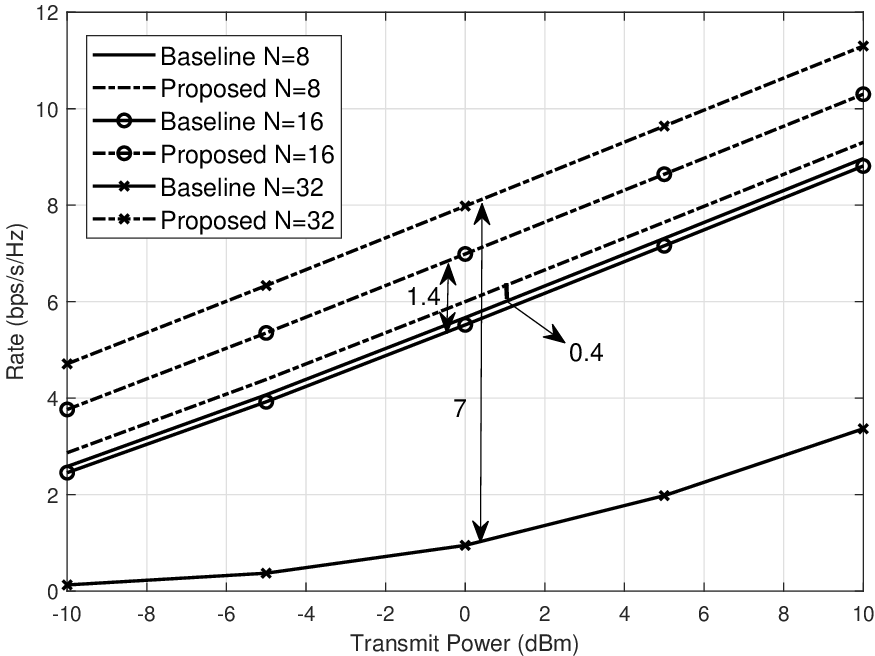}
        \vspace{-3mm}
        \caption{System rate versus transmit power for $M=2$, $d=4$m, $\Delta_u = 0.2$m.}
        \label{fig:diff_pinching_antenna_num}
    \end{minipage} 
    \begin{minipage}{0.32\textwidth}
        \includegraphics[width=0.98\linewidth]{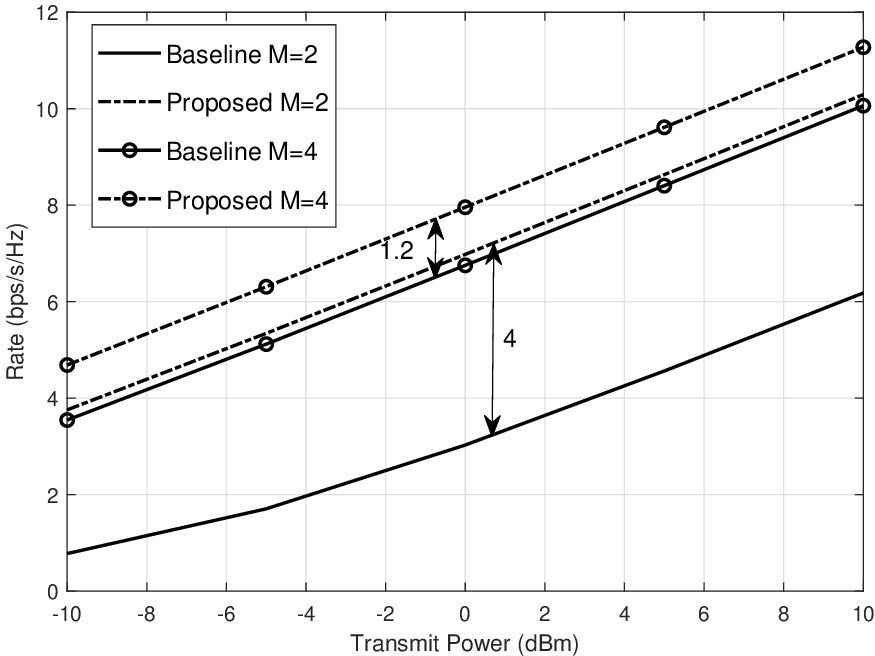}
        \vspace{-3mm}
        \caption{System rate versus transmit power for $N=16$, $d=4$m, $\Delta_u = 0.4$m.}
        \label{fig:diff_ue_antenna_num}
    \end{minipage}
    \begin{minipage}{0.32\textwidth}
        \includegraphics[width=0.98\linewidth]{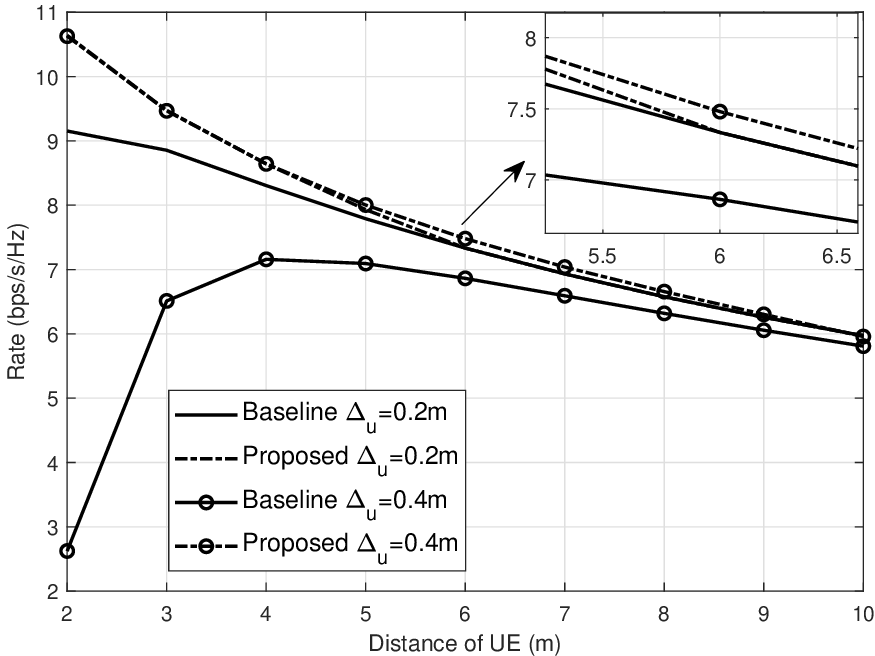}
        \vspace{-3mm}
        \caption{System rate versus distance of user for $N=16$, $M=2$.}
        \label{fig:diff_pinching_antenna_pos}
    \end{minipage}    
    \vspace{-3mm}
\end{figure*}
\setlength{\textfloatsep}{0pt} 

\subsubsection{Algorithm Design}
The proposed antenna placement procedure is carried out in two stages. First, place the $\frac{N}{2}$-th antenna at $x_{\small{{N}/{2}}} = \frac{1}{2}(\tilde{x}_1 + \tilde{x}_M)$ for symmetric layouts or optimize using \eqref{eq:path_loss_objective_function}. 
Once the reference antenna is fixed, the remaining $N-1$ antennas are placed sequentially. Specifically, for each $k = \{2, \ldots, N\}$, the candidate positions $x^o_{m,\pi(k),\left| z \right|}$ are first computed using \eqref{eq:optimal_position_each_antenna}, and then the global position $x^o_{\pi(k)}$ is determined according to Remark 2. Note that in \eqref{eq:optimal_position_each_antenna}, the aggregated channel gain contributed by earlier PA placements is considered through the term $\tilde \Theta _{m,k-1}$.

Specifically, for each receive antenna $m \in \{1,\ldots,M\}$, we define the set of candidate optimal PA positions corresponding to the $m$-th receive antenna as $\mathcal{X}_m = \{x^o_{m,\pi(k),j} \mid j = 1, \ldots, \tilde{Z} \}$, where $\tilde{Z}$ denotes the number of candidate phase-aligned positions. 
The objective is to select one candidate from each $\mathcal{X}_m$ to form a set $\mathcal{S}$ whose spatial span is minimized, as
\begin{equation}
\begin{aligned} 
     & \mathcal{S}^o = \mathop{\arg\min}_{\mathcal{S}} \left\{ \max(\mathcal{S}) - \min(\mathcal{S}) \right\} \\
  &  \text{s.t.} \quad \mathcal{X}_m \cap \mathcal{S} \ne \emptyset, \quad \forall m = 1, \ldots, M,
\end{aligned}
\end{equation}
where $\emptyset$ denotes the empty set.  
\begin{algorithm}[t]
	\caption{  {\color{black}Sliding-window-based Algorithm}   } \label{alg1}
 {\color{black}
	\begin{algorithmic}[1]
		\STATE \textbf{Input}: $\mathcal{X}_m = \{x^o_{m,\pi(k),j}\}_{j=1}^{\tilde Z}$ based on \eqref{eq:optimal_position_each_antenna}; sliding window pointers $p_L = 1$, $p_R = 1$; set window length $W_{\min}=\infty$\;
        \STATE \textbf{Output}: Optimal position subset $\mathcal{S}^o$;
		\STATE \textbf{Sort} all elements $\{x_i\}_{i=1}^{M\tilde Z}$ in ascending order ;
		\WHILE{$p_R \leq M\tilde Z$}
		\STATE   Expand $p_R$ until $\mathcal{X}_m \cap {x_{p_L:p_R}} \ne \emptyset, \forall m$ \; 
	  \WHILE{$\mathcal{X}_m \cap {x_{p_L:p_R}} \ne \emptyset, \forall m$}
        \IF{$p_R-p_L< W_{\min}$}
           \STATE record $p_L^o = p_L$ and $p_R^o = p_R$; 
           \STATE update $W_{\min}=p_R - p_L$;
        \ENDIF
        \STATE  \textbf{update} $p_L = p_L +1$;
		\ENDWHILE 		
		\ENDWHILE 		
        \STATE  \textbf{Return} $\mathcal{S}^o = \{ x_{{p_L^o}:{p_R^o}} \}$ ;  
	\end{algorithmic} }\label{naive_alg}
\end{algorithm}
The $\mathcal{S}^o$ can be found via a sliding-window-based algorithm, as presented in Algorithm \ref{alg1}, it first sorts all candidate phase-aligned positions from each receive antenna, then applies a sliding-window search to identify the most compact cluster covering all $M$ antennas.
The window is iteratively expanded until it contains at least one position from each set 
$\mathcal{X}_m$, and then contracted to minimize the spatial range. 
Then the algorithm identifies a candidate set $\mathcal{S}^o$ that contains at least one position for each receive antenna and has the smallest spatial span, thereby promoting a high composite channel gain across antennas. The $\pi\{k\}$-th PA is then placed at the midpoint of this range, $\mathcal{S}^o$, i.e., $ x^o_{\pi(k)} = \frac{1}{2}(\max\{\mathcal{S}^o\} + \min\{\mathcal{S}^o\})$, which approximately aligns the phases across antennas and yields a near-maximal aggregate channel gain.

After fixing the centrally placed reference antenna, the sliding-window–based algorithm is executed $(N-1)$ times to determine the remaining placements, i.e., $\pi\{ 2\},\ldots,\pi\{ N\}$. 
Each execution requires sorting $M\tilde{Z}$ candidate positions with complexity $\mathcal{O}(M\tilde{Z} \log(M\tilde{Z}))$ and performing a subsequent sliding‐window scan in $\mathcal{O}(M\tilde{Z})$. Therefore, the overall complexity is $\mathcal{O}(NM\tilde{Z} \log(M\tilde{Z}))$, which grows linearly with the number of antennas $N$ and receive elements $M$, and only logarithmically with the candidate set size $\tilde{Z}$.

\vspace{-0.3cm}

\section{Numerical Results and Discussions}
This section evaluates the performance of the proposed antenna placement algorithm. As a benchmark, we consider the scheme from \cite{lv2025beamtrainingpinchingantennasystems} and \cite{10896748}, originally designed for the single-antenna user scenario. To adapt it to the multi-antenna setting, we introduce a virtual user antenna located at the geometric center of the actual array, i.e., $\tilde x_c = 0.5(\tilde x_1 + \tilde x_M)$. The benchmark scheme places the $\frac{N}{2}$-th PA at $\tilde x_c$, then sequentially deploys the remaining antennas by enforcing a fixed phase shift of $2\pi$ between adjacent antennas. Specifically, each new antenna is placed so that its induced channel phase differs by $2z\pi$ from its neighboring antenna, with $z = 1$ for $k > \frac{N}{2}$ and $z = -1$ for $k < \frac{N}{2}$.

Fig. \ref{fig:diff_pinching_antenna_num} compares the proposed scheme with the benchmark for $N = \{8, 16,  32\}$,  $M = 2$, and receive antenna spacing $\Delta_u=\tilde x_M -\tilde x_1 = 0.2$m,  $d = 4$m. As $N$ increases, the benchmark suffers from degraded capacity due to randomized phase alignment across the receive antennas, leading to energy cancellation. In contrast, the proposed method consistently improves the capacity with more antennas, yielding increased performance gains over the benchmark.

Fig. \ref{fig:diff_ue_antenna_num} shows the effect of changing the number of user antennas in a configuration with $N=16$, $d=4$m, and receive antenna spacing $\Delta_u=0.4$m. The antenna positions are set to $(9.7, 10.3)$ for $M=2$ and $(9.7, 9.9, 10.1, 10.3)$ for $M=4$. Both the proposed and benchmark schemes show capacity improvement as the number of user antennas increases. However, the performance gain of the proposed scheme decreases from 4bps/Hz to 1.2bps/Hz, indicating that the proposed approach provides more significant benefits when the user antenna array is sparse.

Fig. \ref{fig:diff_pinching_antenna_pos} presents the capacity performance of the PAS under different user distances, with configuration of $N=16$ and $M=2$. For $\Delta_u = 0.4$m, the benchmark capacity initially increases with distance due to reduced phase difference between the user antennas and the central virtual antenna, but later it declines as path loss dominates in the far field.
In contrast, the proposed scheme exhibits a monotonic decrease in capacity with increasing distance, primarily due to path loss. 
When $\Delta_u = 0.2$m, the proposed and benchmark schemes converge at $d = 6$m. For $\Delta_u = 0.4$m, the proposed scheme retains noticeable gains up to $d = 10$m. These results confirm the effectiveness of the proposed placement strategy in short-range scenarios with multiple user antennas.
 \vspace{-0.3cm}
\section{Conclusions and Future Works}
This paper presented a novel modeling and optimization framework for a downlink PAS with multi-antenna users. By establishing an analytical link between SNR and PA positions, we proposed a two-layer strategy that first aligns the central radiation point using large-scale channel characteristics, followed by a heuristic algorithm for phase coordination across receive antennas. A closed-form solution was derived for the single-antenna case. Future work includes extending to multi-user scenarios, incorporating real-time environmental feedback, and enhancing robustness under NLoS conditions.

\vspace{-0.25cm}


\bibliography{bibliography}
\bibliographystyle{IEEEtran}

%
%
%

\end{document}